\documentclass[aps,pre,citeautoscript,twocolumn,reprint,longbibliography,floatfix]{revtex4-2} 

\usepackage[T1]{fontenc}
\usepackage{lmodern}
\usepackage{xcolor}
\usepackage{here}
\usepackage{graphicx}
\usepackage{amssymb}
\usepackage{amsmath}
\usepackage{physics}
\usepackage{algorithm}
\usepackage{algorithmic}
\usepackage{upgreek}
\usepackage{bm}
\usepackage{mathtools}
\usepackage{siunitx}
\usepackage{hyperref}
\usepackage[capitalize]{cleveref}

\newcommand{\etal}{\textit{et al}.\ }

\newcommand*{\citeref}[1]{Ref.~\onlinecite{#1}}
\newcommand*{\citesref}[1]{Refs.~\onlinecite{#1}}

\crefname{fig}{Fig.}{Figs.}
\Crefname{fig}{Figure}{Figures}

\crefname{tab}{Table}{Tables}
\Crefname{tab}{Table}{Tables}

\crefname{alg}{Algorithm}{Algorithms}
\Crefname{alg}{Algorithm}{Algorithms}

\crefname{sec}{Sec.}{Secs.}
\Crefname{sec}{Section}{Sections}

\crefname{eq}{Eq.}{Eqs.}
\Crefname{eq}{Equation}{Equations}

\begin{document}

\title{Stochastic parameter optimization analysis of dynamical quantum critical phenomena in long-range transverse-field Ising chain}

\author{Sora Shiratani}
\email{sora.shiratani@phys.s.u-tokyo.ac.jp}
\affiliation{Department of Physics,\! The University of Tokyo,\! Tokyo 113--0033,\! Japan}

\author{Synge Todo}
\email{wistaria@phys.s.u-tokyo.ac.jp}
\affiliation{Department of Physics,\! The University of Tokyo,\! Tokyo 113--0033,\! Japan}
\affiliation{Institute for Physics of Intelligence,\! The University of Tokyo,\! Tokyo 113--0033,\! Japan}
\affiliation{Institute for Solid State Physics,\! The University of Tokyo,\! Kashiwa,\! 277--8581,\! Japan}

\date{\today}

\begin{abstract}
    The quantum phase transition of the one-dimensional long-range transverse-field Ising model is explored by combining the quantum Monte Carlo method and stochastic parameter optimization, specifically achieved by tuning correlation ratios so that space and imaginary time are isotropic.
    In our simulations, the simulator automatically determines the parameters to sample from, even without prior knowledge of the critical point and universality class.
    The leading-order finite-size corrections are eliminated by comparing two systems with different sizes; this procedure is also performed automatically.
    Varying the decay exponent of the long-range interaction, \(\sigma\), we investigate \(\sigma\)-dependence of the dynamical exponent and the other critical exponents precisely in the mean-field, non-universal, and two-dimensional classical Ising universality regimes.
    We successfully obtained numerical evidence supporting \(\sigma = 7/4\) as the universality boundary between the latter two.
\end{abstract}

\maketitle


\section{Introduction}\label{sec:intro}
Long-range interacting systems have been a subject of interest in statistical physics for decades.
The earliest attempts date back to the 1970s when the modern renormalization group (RG) theory was still being developed~\cite{Nagle1970,Anderson1971,Fisher1972,Sak1973,Suzuki1973critical,Yamazaki1977}.
At that time, they were mainly quoted as a playground for theoreticians.
In the 21st century, however, the situation has changed drastically~\cite{Defenu2023}; we now have access to more sophisticated theories, better experimental techniques, and more powerful computers.
Fueled by these advances, long-range interacting systems are growing into a new field of research covering theoretical, experimental, and numerical studies.
From the theoretical point of view, for the reputed RG approach~\cite{Yamazaki1977, Thouless1969, Suzuki1973critical, Sak1973, Nagle1970, Lipkin1965, Fisher1972, Cannas1995, Anderson1971, Dutta2001, Defenu2017}, they are also attracting attention in the context of Kibble-Zurek scaling~\cite{Jaschke2017, Dutta2017, Puebla2019, Li2023, Puebla2020} and nonequilibrium physics~\cite{Halimeh2017, Lang2018dynamical, Lang2018concurrence, Sperstad2012}.
Experiments are also being carried out in various fields, including spin-ice materials~\cite{Steven2001, Castelnovo2008}, quantum optics~\cite{Britton2012, Islam2013, Jurcevic2014, Richerme2014, Justin2016, Yang2019, Monroe2021}, Rydberg atoms~\cite{Saffman2010, Schauss2012, Labuhn2016, Landig2016, Browaeys2020}, etc.
Numerical techniques have also developed in parallel, such as tensor network~\cite{Vodola2016, Vanderstraeten2018, Zhu2018, White1992, Crosswhite2008, Pirvu2010} and Monte Carlo methods for long-range interacting systems~\cite{Fukui2009, Lujiten2002, Humeniuk2020, Humeniuk2016, Katzgraber2003, Koziol2021, Sperstad2012, Sperstad2010, Tomita2009, Sandvik2003, Eduardo2021,Michel2019, Muller2023}.

Among many rich properties of long-range interacting systems, quantum phase transition (QPT) under strong space-time anisotropy would be one of the most intriguing and challenging phenomena.
Be it long-range or not, QPT, phase transition at zero temperature, is often associated with competing interactions.
The ferromagnetic long-range transverse-field Ising model (LRTFIM), which is the main subject of the present paper, is a typical example of systems with competition.
It has the following Hamiltonian:
\begin{equation}
    \begin{split}
        H & = -\sum_{i < j} J_{ij} Z_i Z_j - \Gamma \sum_i X_i                              \\
          & = -\sum_{i < j} \frac{1}{\qty|i - j|^{1 + \sigma}} Z_i Z_j - \Gamma \sum_i X_i,
    \end{split}
    \label{eq:model}
\end{equation}
where \(X_i\) and \(Z_i\) are Pauli operators acting on the \(i\)-th spin on one-dimensional (1D) chain, \(\Gamma\) denotes the strength of the transverse field, and most importantly, \(\sigma > 0\) is the decay exponent of the ferromagnetic long-range interaction.
We assume \(\Gamma \geq 0\) without loss of generality as we can change the sign of \(\Gamma\) by unitary transformation \(\qty(X,Z) \mapsto \qty(-X,-Z)\).

We can readily identify two conflicting parts in \cref{eq:model}: the long-range Ising interaction favoring ferromagnetic order along the spin \(z\) axis and the transverse-field term making spins point to the \(x\) direction.
Consequently, by adjusting \(\Gamma\), we can trigger QPT from a two-fold-degenerating ferromagnetic phase to a disordered one at a certain critical point \(\Gamma_\text{c}\).
For LRTFIM, in addition to \(\Gamma\), we have another parameter \(\sigma\).
In contrast to \(\Gamma\), which directly controls the competition between the two terms, \(\sigma\) affects the critical property of QPT\@.
In the \(\sigma = \infty\) limit, the model reduces to the nearest-neighbor transverse-field Ising model, which can be transformed further into the two-dimensional (2D) classical nearest-neighbor Ising model with anisotropic coupling strength~\cite{Lieb1961, Sachdev2011}.
The transverse-field term is mapped to the nearest-neighbor interaction along the imaginary-time \(\tau\) axis~\cite{Lieb1961, Suzuki1976, Sachdev2011, Gubernatis2016}.
On the other hand, for small \(\sigma\), as each spin has an extensive number of ``neighbors'' due to the slow decay of the interaction, the system is expected to exhibit mean-field-like behavior.
Thus, we can recognize \(\sigma\) as a parameter that controls \textit{the effective dimension} of the system from 2D (\(\sigma = \infty\)) to infinity (\(\sigma = 0\)).

With another parameter \(\sigma\) in hand, it is natural to ask how the critical property of QPT changes as we vary \(\sigma\).
As seen later, the mean-field universality is observed for small \(\sigma\).
With the RG theory, we can compute the critical exponents and the upper boundary of the mean-field region, which is \(\sigma = 2 / 3\) in the 1D case\@.
In contrast, the situation is a bit more complicated for larger \(\sigma\).
Although we can readily convince ourselves that the 2D classical Ising universality appears for \(\sigma \geq 2\), it is not easy to nail down its lower boundary.
As discussed later, there are two conflicting predictions currently; one is \(\sigma = 2\)~\cite{Fisher1972, Zhu2018, Dutta2001, Knap2013, Suzuki1973calculation, Suzuki1973critical, Yamazaki1977, Picco2012, Blanchard2013, Grassberger2013} and the other is \(\sigma = 7 / 4\)~\cite{Sak1973, Gori2017, Lujiten2002, Angelini2014, Defenu2017, Horita2017, Paulos2016, Behan2017scaling, Behan2017longrange, Behan2019}.
Finally, there is a non-universal region between these two universalities, where the critical exponents depend on \(\sigma\) in a nontrivial manner.
In this region, numerical simulations are the most promising approach since long-range interaction strongly governs the system.

Let us briefly review the current status from a numerical point of view.
When one wants to simulate beyond small systems handleable by exact diagonalization~\cite{Knap2013, Sandvik2010, Hauke2015, Li2016, Homrighausen2017, Zhu2018, Jaschke2017, Frowis2010, Murg2010}, the most potent methods would be tensor network and the quantum Monte Carlo (QMC).
Initially, both methods were developed for short-range interacting systems, but later, they have been extended to long-range interacting systems.
For the former, with the matrix product operator formulation~\cite{Crosswhite2008, Pirvu2010}, we can now perform tensor network simulations as if there were only short-range interactions~\cite{Vodola2016, Vanderstraeten2018, Zhu2018}.
Within this framework, Jaschke \etal~\cite{Jaschke2017} estimated \(\Gamma_\text{c}\) across a wide range of \(\sigma\) using iMPS and finite-size scaling (FSS), and Puebla \etal~\cite{Puebla2019} did for several critical exponents in addition to \(\Gamma_\text{c}\).
Nonetheless, from its formulation based on the low-rank approximation of tensors, tensor network simulations always entail nontrivial systematic errors.
In contrast, QMC is free from such systematic errors; using the modern continuous-time formulation~\cite{Fukui2009, Gubernatis2016}, we can realize the exact canonical ensemble governed by the Hamiltonian \cref{eq:model}.
Additionally, for LRTFIM, we are not bothered by the negative sign problem~\cite{Gubernatis2016}, a notorious problem in QMC simulations.
With these advantages, QMC simulations have been one of the most powerful tools to investigate LRTFIM\@.
Limiting the scope to the most recent studies, Koziol \etal~\cite{Koziol2021} and Eduardo \etal~\cite{Eduardo2021} applied QMC and FSS to estimate the critical exponents.

Indeed, the state-of-the-art QMC algorithm, the order-N continuous-imaginary-time cluster algorithm~\cite{Fukui2009}, works almost in an optimal way for evaluating the thermal average of physical quantities of LRTFIM for a given parameter set, \(\qty(\sigma,L,K,\Gamma)\), where \(L\) is the system size and \(K = 1 / T\) is the inverse temperature.
Even so, precise FSS analysis for LRTFIM is still challenging.
This is because of strong space-time anisotropy in the presence of the long-range interaction only in the spatial direction.
The strength of the space-time anisotropy at QPT is characterized by the dynamical exponent \(z\), which is unity in the 2D classical Ising regime but can take a nontrivial value for small \(\sigma\).
Consequently, as we see later, in the presence of space-time anisotropy, we have two independent FSS fields, \(L^{z} / K\) and \(L^{1 / \nu} \qty(\Gamma - \Gamma_\text{c})\), in the FSS form, being \(\nu\) the critical exponent for the spatial correlation length.
The existence of two independent fields means that for each \(L\), we have to perform a wasting search on a dense grid in the \((K, \Gamma)\) plane for precise FSS analysis.
Such requirement of massive computational cost has prevented previous QMC studies for detailed analysis around the most challenging region, \(\sigma \simeq 7 / 4\), and exhaustive analysis over a wide range of \(\sigma\) so far.

\begin{figure}[tbp]
    \centering
    \includegraphics[width=0.8\linewidth]{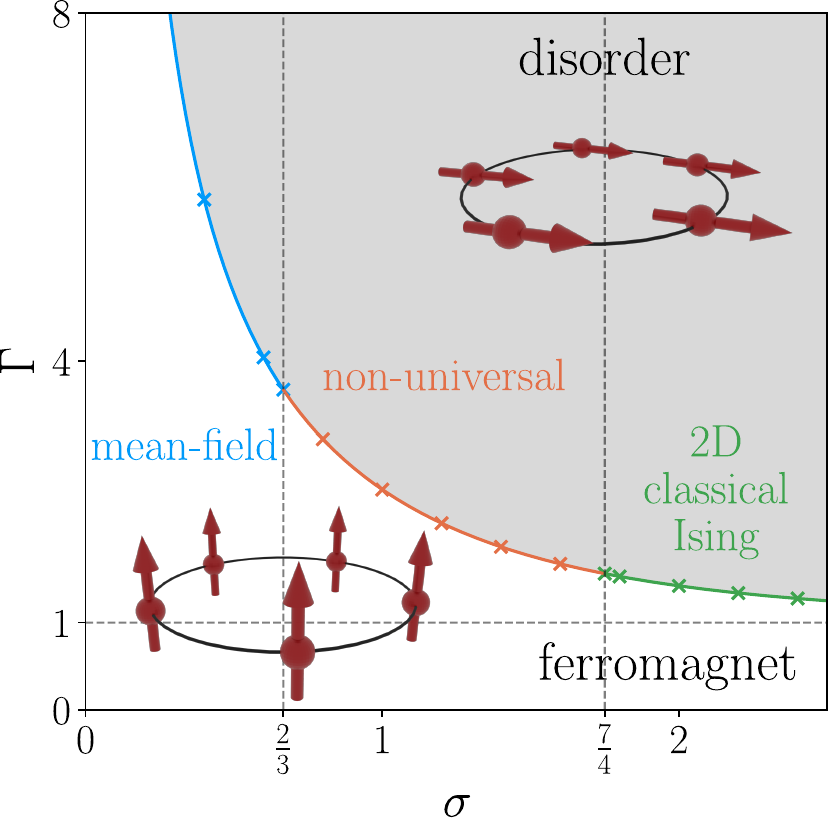}
    \caption{%
        Ground state phase diagram of 1D LRTFIM\@.
        The crosses represent our numerical results.
        The solid curve represents the empirical formula [\cref{eq:empirical} in \cref{sec:result}], which effectively captures the expected asymptotic behavior for both \(\sigma \to 0\) and \(\sigma \to \infty\), and consistently describes the present numerical results within 95\% confidence interval across the entire range of \(\sigma\).
        For \(\sigma \leq 2 / 3\), \(2 / 3 \leq \sigma \leq 7 / 4\), and \(7 / 4 \leq \sigma\), QPT is classified into the mean-field, non-universal, and 2D classical Ising universality classes, respectively.
        The horizontal dotted line denotes the exact critical transverse field, \(\Gamma_\text{c} = 1\), in the \(\sigma = \infty\) limit.
    }\label{fig:diagram}
\end{figure}

In this paper, we analyze LRTFIM by a new approach that combines the order-N QMC~\cite{Fukui2009} with stochastic parameter optimization~\cite{Robbins1951, Yasuda2015}.
In our simulations, only the decay exponent \(\sigma\) and the system size \(L\) are the input parameters, and the inverse temperature \(K\) and the transverse field \(\Gamma\) are not; \(K\) and \(\Gamma\) are automatically optimized so that the system is at the finite-size quantum critical point with the isotropic space-time geometry.
This procedure forces observables to obey a power law regarding system size \(L\), and scaling powers are directly linked to critical exponents~\cite{Sachdev2011, Yasuda2015, Sperstad2012}.
We can extract critical properties from a set of simulations with different \(L\)'s, enabling us to perform an exhaustive analysis of the phase diagram without a wasting search on a dense grid in the \(\qty(K, \Gamma)\) plane.
In our proposed method, the hyperparameter in the original algorithm~\cite{Yasuda2015} is also optimized automatically; it is determined so that the optimized value of \(\Gamma\) achieves the fastest convergence to the critical point as \(L\) increases, aiming at mitigating the finite-size effect introduced by the shift from the critical point.
It enables us to evaluate the critical exponents, including \(z\), with high accuracy and determine the universality boundaries precisely.

Let us close this section by showing the ground state phase diagram in advance.
In \cref{fig:diagram}, the solid curve represents the phase boundary between the ferromagnetic and disorder phases on the \((\sigma, \Gamma)\) plane.
The critical transverse field \(\Gamma_\text{c}\) decreases monotonically as \(\sigma\) increases and tends to converge to the exact value, \(\Gamma_\text{c} = 1\), in the \(\sigma = \infty\) limit~\cite{Lieb1961, Gubernatis2016}.

The rest of the present paper is organized as follows:
In \cref{sec:theory}, we survey previous theoretical analyses on LRTFIM, focusing on the universality class of QPT\@.
In \cref{sec:method}, we introduce our scheme to combine QMC with stochastic parameter optimization~\cite{Robbins1951, Yasuda2015} to achieve efficiency and accuracy with a minimum cost: this is the main contribution of this paper.
We present our numerical results in \cref{sec:result}.
Finally, in \cref{sec:discussion}, we discuss our results and conclude this paper.


\section{Theoretical background}\label{sec:theory}

\subsection{Theoretical predictions}
To begin with, let us briefly review the theoretical predictions for QPT of LRTFIM\@.

Several observables show distinctive power-low behavior near quantum critical points, and associated critical exponents are defined~\cite{Sachdev2011}.
As QPT occurs at absolute zero, the degrees of singularity are measured using a control parameter \(g\) rather than shifted temperature.
For LRTFIM, \(g\) is shifted transverse field \(g = \Gamma - \Gamma_\text{c}\).
Aside from this difference, definitions are pretty similar to their classical counterparts; \(\upbeta\) is defined from squared magnetization \(m^2 \sim g^{2\upbeta}\), \(\upgamma\) from susceptibility \(\chi \sim 1 / g^\upgamma\) and \(\nu\) from correlation length \(\xi_x \sim 1 / g^\nu\).
Additionally, the dynamical exponent \(z\), which is specific to QPT, is defined from correlation length along the imaginary-time axis \(\xi_\tau \sim 1 / g^{z \nu}\).

Let us consider the continuum description of LRTFIM as a starting point for the RG analysis.
The action for LRTFIM is the same as the \((1 + 1)\)-dimensional \(\phi^4\) model up to an additional term \(\propto |q|^\sigma\) accounting for the long-range interactions~\cite{Dutta2001, Defenu2017}:
\begin{equation}
    -\int \frac{\dd{q}}{2\pi} \frac{\dd{\omega}}{2\pi} \tilde{\phi}^*\qty(g \omega^2 + r + c\qty|q|^{\sigma} + dq^2)\tilde{\phi} - \int \dd{x} \dd{\tau} \frac{u}{4!}\phi^4 \label{eq:action},
\end{equation}
where \(\qty(x,\tau)\) denote the space and imaginary time, \(\qty(q,\omega)\) the corresponding wavenumbers, \(\phi\) is a scalar field at \((x,\tau)\), \(\tilde{\phi}\) is the Fourier transform of \(\phi\), and \(\qty(g,r,c,d,u)\) are the coupling constants; here, only \(r \propto \Gamma - \Gamma_\text{c}\) can be negative, and QPT occurs at \(r = 0\)~\cite{Sachdev2011}.
We can naturally extend the action to general \(\qty(D + 1)\)-dimensional cases.

First, note that the long-range nature of the system is encoded in the \(\qty|q|^\sigma\) term, which competes with the \(q^2\) term derived from short-range interactions.
As small wavenumbers govern critical phenomena, we can naturally expect short-range behavior for \(\sigma > 2\).
For the 1D case, the short-range limit (\(\sigma = \infty\)) belongs to the 2D classical Ising universality class~\cite{Lieb1961, Gubernatis2016}.
Similarly, we can dig down the opposite limit, i.e., ultra-long-range limit \(\sigma \simeq 0\), where the mean-field theory describes the critical behavior.
Here, same as the \(\phi^4\) model, ``mean-field'' stands for the universality class linked to the Gaussian fixed point \(\qty(r,u) = \qty(0,0)\)~\cite{Sachdev2011}.
In this regime, we can readily perform diagram expansion as the \(\phi^4\) term is diagonal in the Fourier space.
The result is also very similar to the \(\phi^4\) model, with its upper critical dimension, \(4\), replaced by \(D_u = 2\sigma - z = 3 \sigma / 2\)~\cite{Dutta2001, Defenu2017, Sachdev2011}.
Therefore, in the 1D case, the mean-field universality is expected for \(\sigma \le 2 / 3\).
In this regime, we can also evaluate the critical exponents analytically as listed in \cref{tab:exp}.

At a glance, it seems reasonable to conclude that \(\sigma = 2\) is the lower boundary of the Ising universality.
However, the situation could not be so straightforward.
The point is that the scaling exponent of \(c\) is exactly given by \(y_c = 2 - \sigma - \upeta\)~\cite{Defenu2017, Sachdev2011}, where \(\upeta\) is the anomalous dimension of \(\phi\).
Apparently, \(y_c|_{\sigma = 2} = - \upeta_{\textrm{Ising}} = - 1 / 4\) could be too small to be the lower boundary; requiring \(y_c = 0\), we can identify \(\sigma = 7 / 4\) as another candidate.
This naive argument can further be refined to cover higher-order corrections~\cite{Sak1973}, leading to the same conclusion.
Indeed, this \(2\) vs \(7 / 4\) dispute has been a long-standing problem~\cite{Fisher1972, Zhu2018, Dutta2001, Knap2013, Suzuki1973calculation, Suzuki1973critical, Yamazaki1977, Picco2012, Blanchard2013, Grassberger2013, Sak1973, Gori2017, Lujiten2002, Angelini2014, Defenu2017, Horita2017, Paulos2016, Behan2017scaling, Behan2017longrange, Behan2019}.
This paper also aims to settle this issue; as we will see later, our simulation supports the latter.

For the intermediate regime, \(2 / 3 < \sigma < 7 / 4\), we have to deal with the long-range nature of the system directly.
At present, analyses are mainly based on numerical approaches such as series expansion~\cite{Fey2016, Adelhardt2020}, exact diagonalization~\cite{Knap2013, Sandvik2010, Hauke2015, Li2016, Homrighausen2017, Zhu2018, Jaschke2017, Frowis2010, Murg2010}, tensor network~\cite{Vodola2016, Vanderstraeten2018, Zhu2018}, functional RG~\cite{Defenu2017} and QMC~\cite{Koziol2021, Sandvik2003, Eduardo2021, Sperstad2010, Sperstad2012, Humeniuk2016, Humeniuk2020}.

Although it is out of the scope of this paper, let us note that for \(\sigma \le 1\), a finite-temperature phase transition exists in addition to the QPT we are focusing on.
Interestingly, the finite-temperature ordered phase abruptly disappears at \(\sigma = 1\) via the Kosterlitz-Thouless phase transition~\cite{Dutta2001, Anderson1971, Cannas1995, Luijten2001, Sandvik2003, Fukui2009, Humeniuk2020}.

\begin{table}[tbp]
    \centering
    \begin{ruledtabular}
        \begin{tabular}{ccc}
            \mbox{\hspace*{1cm}} & mean-field             & 2D Ising  \\ \hline 
            \(\upbeta\)          & \(1 / 2 - \sigma / 4\) & \(1 / 8\) \\
            \(\upgamma\)         & \(1\)                  & \(7 / 4\) \\
            \(\nu\)              & \(1 / \sigma\)         & \(1\)     \\
            \(z\)                & \(\sigma / 2\)         & \(1\)     \\
        \end{tabular}
    \end{ruledtabular}
    \caption{Critical exponents, \(\upbeta\), \(\upgamma\), \(\nu\), and \(z\), in the mean-field regime (\(\sigma \leq 2 / 3\)) and the 2D Ising regime (\(\sigma \geq 7 / 4\)).}\label{tab:exp}
\end{table}

\subsection{Finite-size scaling}\label{sec:fss}
The FSS theory is a powerful tool to bridge numerical results for finite-size systems and the critical phenomena in the thermodynamic limit.
According to the RG theory~\cite{Sachdev2011}, the singular part of free energy density is transformed as
\begin{equation}
    f_{\mathrm{s}}\qty(L,g_1,g_2,\dots) \sim b^{-y_f} f_{\mathrm{s}}\qty(L / b,b^{y_1} g_1, b^{y_2} g_2,\dots), \label{eq:fssb}
\end{equation}
where \(g_\ast\) are the model parameters measured from the critical point, \(y_\ast\) are the scaling exponents, and \(b\) is the RG scaling factor.

For LRTFIM, we have two types of FSS depending on \(\sigma\):
First, for \(\sigma \geq 2 / 3\), \cref{eq:fssb} directly applies for relevant and marginal parameters, \(g_1 = 1/K\), \(g_2 = g = \Gamma - \Gamma_\text{c}\), and \(g_3 = h\) (longitudinal field conjugate to \(\sum_i Z_i\)):
\begin{equation}
    f_{\mathrm{s}} \sim b^{-(1 + z)} f_{\mathrm{s}}\qty(L / b,b^{y_t} / K,b^{y_g} g,b^{y_h} h). \label{eq:fsstfi}
\end{equation}
In contrast, for \(\sigma < 2 / 3\), we have to be careful for \(u\), which is the coupling constant of \(\phi^4\) term in \cref{eq:action}.
This term is dangerously irrelevant in this regime; the free energy diverges near the critical point \(u = 0\) due to the absence of the quartic term.
To incorporate this effect, we further assume \(\mathcal{F}\qty(t,g,h,u) \sim u^{p_u} \mathcal{F}\qty(u^{p_t} t,u^{p_g} g,u^{p_h} h)\) near \(u = 0\), where \(p_\ast\) absorb the divergence~\cite{Binder1985}.
Finally, we get
\begin{equation}
    f_{\mathrm{s}} \sim b^{-(1 + z)} f_{\mathrm{s}}\qty(L / b,b^{y_t + p_t y_u} / K,b^{y_g + p_g y_u} g,b^{y_h + p_h y_u} h).\label{eq:fsstfi2}
\end{equation}
Due to \(p_\ast\), while the appearance of \cref{eq:fssb} itself is still valid, all the exponents are modified, and thus their meanings are obscured; they are no longer the scaling exponents of the original parameters.
Anything derived from \(f\), such as squared magnetization, inherits this problem.
Even still, \(y_f = 1 + z\) seems valid due to its direct correspondence to physicality, as argued for classical systems~\cite{Binder1985}.
Later, we will see that our method can bypass this problem for some critical exponents.

From the above discussions, we reach the FSS ansatz for the free energy density by setting \(b = L\):
\begin{equation}
    f_{\mathrm{s}} \sim L^{-(1 + z)} \mathcal{F}\qty(L^{y_t}/K,L^{y_g} g). \label{eq:fsspre}
\end{equation}
Here, we dropped \(h\) because we consider the \(h = 0\) case.
Due to the above-mentioned problem, for \(\sigma < 2 / 3\), \(y_\ast\) no longer have their original meanings.
Optionally, we can rewrite \(y_\ast\) by more intuitive exponents via \(\xi_x \sim 1 / g^\nu\) and \(\xi_\tau \sim 1 / g^{z \nu}\).
Finally, we get
\begin{equation}
    f_{\mathrm{s}} \sim L^{-(1 + z)} \mathcal{F}\qty(L^{z} / K,L^{1 / \nu} g) \label{eq:fss}
\end{equation}
and more generally for any quantity \(Q\),
\begin{equation}
    Q \sim L^{-x_Q} \mathcal{Q}\qty(L^{z} / K,L^{1 / \nu} g), \label{eq:fssq}
\end{equation}
where \(x_Q\) is the scaling dimension of \(Q\).


\section{Method}\label{sec:method}
\subsection{Quantum Monte Carlo method}
As a building block of our framework, we employ QMC to evaluate thermal averages of physical quantities.
In this paper, we adopt the order-N continuous-imaginary-time cluster algorithm~\cite{Fukui2009}.

The starting point of the algorithm is the Swendsen-Wang cluster algorithm for the classical Ising model~\cite{Swendsen1987}.
This algorithm creates spin clusters by stochastically bonding interacting spin pairs and then flips spins in each cluster collectively.
The autocorrelation time of observables is typically much shorter than that of conventional local update algorithms.
Utilizing the continuous-time path integral formulation, we can extend this method to the transverse field Ising model without any Suzuki-Trotter discretization error or cutoff in the expansion of operator exponential~\cite{Gubernatis2016,Beard1996,Blote2002}.
The algorithm's time complexity per sweep is proportional to the total number of interacting pairs and the inverse temperature.

Extending the cluster algorithm further to the models with long-range interactions is straightforward.
The sampling process remains almost the same, the only difference being that cluster formation is not restricted to nearest neighbors.
However, due to LRTFIM having \(\binom{L}{2} = \order{L^2}\) interacting spin pairs, the time complexity per sweep increases proportionally to \(L^2\), instead of \(L\), in a naive implementation.

To avoid such enormous sampling costs, we focus on the fact that in the continuous-imaginary-time cluster algorithm, the bonding success probability is proportional to \(J\).
In the order-N cluster algorithm~\cite{Fukui2009}, we first choose a bond according to the above probability and then check if the bonding at that position can be accepted instead of inspecting each bond in turn.
This way, the computational complexity per sweep of the modified procedure becomes \(\order{L}\) since the average number of cluster bonds is \(\order{L}\) as long as the ground state energy of the system is extensive, i.e., \(\sigma > 0\) in \cref{eq:model}.
Note that this order-N algorithm handles long-range interaction without any approximations, such as introducing a finite cutoﬀ.
See \citesref{Fukui2009,TodoS2013} for more details.

\subsection{Finite-size scaling without data collapse}
Usually, the FSS ansatz is mainly used as the fitting function for simulation data.
Regarding this conventional treatment, let us point out two problems.
First, there is no prior knowledge for \(\mathcal{Q}\) and \(\Gamma_\text{c}\).
This means that we more or less have to do try-and-errors to estimate them roughly at the first step.
Second, we have two independent FSS fields, so we must perform a wasting search on a dense grid in the \((K, \Gamma)\) plane for each \(\sigma\) and \(L\) for precise FSS analysis.

To overcome these problems, we introduce a new approach based on stochastic approximation~\cite{Yasuda2015}.
To begin with, consider the following FSS ansatzes for two correlation lengths \(\xi_x\) and \(\xi_\tau\):
\begin{align}
    \xi_x / L    & \sim \Xi_x\qty(L^{z} / K,L^{1 / \nu} g), \label{eq:fssx}    \\
    \xi_\tau / K & \sim \Xi_\tau\qty(L^{z} / K,L^{1 / \nu} g). \label{eq:fsst}
\end{align}
Here, we assume that their scaling exponents vanish after rescaling by corresponding lengths, which is justified by RG theory.
As previously mentioned, by writing down these formulae, we implicitly assume that \(\sigma \ge 2 / 3\).
Next, we adjust \(\qty(K,g)\) so that \(\xi_x / L = \xi_\tau / K = R > 0\) is satisfied.
Physically speaking, this implies that we force the system to be ``isotropic.''
Here, we leave \(R\) as a free parameter to be determined later.
Provided we somehow succeeded in this optimization, we get \(\Xi_x\qty(L^{z} / K ,L^{1 / \nu} g) = \Xi_\tau\qty(L^{z} / K,L^{1 / \nu} g) = R\), which defines two constraints for two unknowns \(\qty(L^{z} / K,L^{1 / \nu} g)\).
Comparing degrees of freedom, we should get \(\qty(L^{z} / K,L^{1 / \nu} g) = \textrm{const}\)., simplified to \(K \propto L^{z}\) and \(g = \Gamma - \Gamma_\text{c} \propto L^{-1 / \nu}\).
Here, we arrive at an important property; \(\Gamma_\text{c}\) and critical exponents are drawn from \(L\)-dependence of \(K\) and \(\Gamma\).
This statement also applies to general \(Q\), directly obtained from \(\qty(L^{z} / K,L^{1 / \nu} g) = \textrm{const}\).
In general, as \(\mathcal{Q}\qty(L^{z} / K,L^{1 / \nu} g) = \textrm{const}\)., we have \(Q \sim L^{-x_Q}\).

Interestingly, the above property is partially true even for \(\sigma < 2 / 3\).
From the definition of \(\nu\) and \(z\), we have \(\xi_x \sim 1 / g^{\nu}\) and \(\xi_\tau \sim 1 / g^{z \nu}\) regardless of \(\sigma\).
As they are both coerced to be equal to \(R\) after rescaling, we get \(1 / g^{\nu} \sim LR\) and \(1 / g^{z \nu} \sim K R\).
Erasing \(g\), we again get \(K \propto L^{z}\) and \(\Gamma - \Gamma_\text{c} \propto L^{-1 / \nu}\) even though \cref{eq:fssx,eq:fsst} can no longer be valid.
This counterintuitive result is supported by the fact that \(\xi_x / L\) and \(\xi_\tau / K\) can be finite even with dangerously irrelevant variables~\cite{Binder1985}.
Unfortunately, as we still have \(p_\ast\) in \cref{eq:fsstfi2} even though \(K \propto L^{z}\) and \(\Gamma - \Gamma_\text{c} \propto L^{-1 / \nu}\), \(L\)-dependence of \(Q\) does not provide direct information on \(x_Q\).

\subsection{Stochastic approximation}
We use stochastic approximation as the concrete method for solving the equation systems.
Let us consider a model problem finding the unique root of a \(\mathbb{R} \to \mathbb{R}\) monotonic function \(f\qty(x)\) from its noisy independent samples \(f\qty(x) + \varepsilon\)~\cite{Robbins1951, Bishop2013}, where \(\varepsilon\) can be anything as long as it is i.i.d.\ (independent and identically distributed) and has zero mean.
The stochastic approximation method solves this problem by the following iterative procedure:
\begin{algorithm}[H]
    \begin{algorithmic}[1]
        \REQUIRE{Current estimate \(x_n\), hyperparameter \(A\)}
        \STATE{Sample \(f_n \sim f\qty(x_n) + \varepsilon\)}
        \STATE{\(x_{n + 1} \leftarrow x_n + A f_n / n\)}
        \ENSURE{\(x_{n + 1}\)}
    \end{algorithmic}
    \caption{Stochastic approximation for \(f\qty(x)= 0\).}\label{alg:sa}
\end{algorithm}
Here, \(A\) is a hyperparameter significantly affecting the convergence speed~\cite{Yasuda2015}.
As intuitively expected, \(A\) should be negative for increasing \(f\) and positive for the opposite.
For multidimensional problems, we can extend \cref{alg:sa} by replacing \(A\) with real matrix \(\boldsymbol{A}\) and \(x\) with real vector \(\boldsymbol{x}\)~\cite{Blum1954}.
The target function is also extended to \(\bm{f}\qty(\boldsymbol{x}) = \bm{0}\), defining the same number of constraints with the dimension of \(\boldsymbol{x}\).
In the present simulations, \(\boldsymbol{x}\) and \(\bm{f}\) correspond to model parameters \((K,\Gamma)\) and constraints written in terms of correlation ratios, respectively.
Theoretically, optimal choice of \(\boldsymbol{A}\) is \(\boldsymbol{A} \simeq -\qty[\left.\grad{\bm{f}}\right|_{\bm{f} = \bm{0}}]^{-1}\)(see \citesref{Yasuda2015, Chung1954}).
However, estimating \(\boldsymbol{A}\) accurately before simulation is not straightforward.
In the present study, instead of optimizing \(\boldsymbol{A}\), we construct a linear combination of equations so that \(\boldsymbol{A}\) is expected to be nearly diagonal, based on physical insights, as explained in detail in the next section.

\subsection{Stochastic parameter optimization}

\subsubsection{Finding critical point and optimization of aspect ratio}

We use the second moment of the correlation functions~\cite{Cooper1982,Todo2001,Suwa2015} to numerically estimate the correlation lengths \(\xi_x\) and \(\xi_\tau\):
\begin{align}
    \frac{\xi_x}{L}    & = \frac{1}{2 \pi} \sqrt{\frac{S(0,0)}{S(\frac{2\pi}{L},0)} - 1}, \label{eq:xi_x} \\
    \frac{\xi_\tau}{K} & = \frac{1}{2 \pi} \sqrt{\frac{S(0,0)}{S(0,\frac{2\pi}{K})} - 1}, \label{eq:xi_t}
\end{align}
where \(S\qty(q,\omega)\) is the dynamical structure factor at momentum \(q\) and Matsubara frequency \(\omega\):
\begin{align}
    S\qty(q,\omega) =  \frac{1}{K L} \langle \Big|\sum_{x} e^{iqx} \int \dd{\tau} e^{i \omega \tau} Z(x,\tau) \Big|^2 \rangle.
    \label{eq:dsf}
\end{align}
Note that while \cref{eq:dsf} is an unbiased estimator for \(S\qty(q,\omega)\), \cref{eq:xi_x,eq:xi_t} are not as they involve nonlinear operations of observables: quotient and square root.
Such bias becomes especially prominent in the present method, as we use the average of only a few Monte Carlo steps (MCS) at each iteration of stochastic approximation (\(f_n\) in \cref{alg:sa}).
To deal with this problem, we transform \(\xi_x/L=R\) and \(\xi_\tau/K=R\) into the following forms, which do not include nonlinear transformations:
\begin{gather}
    S\qty(0,0) - \qty(1 + 4 \pi^2 R^2) S\qty(\frac{2\pi}{L},0) = 0,  \label{eq:tuneb_naive} \\
    S\qty(0,0) - \qty(1 + 4 \pi^2 R^2) S\qty(0,\frac{2\pi}{K}) = 0. \label{eq:tuneg_naive}
\end{gather}
This time, the left-hand sides of \cref{eq:tuneb_naive,eq:tuneg_naive} contain only the unbiased estimators~\cite{Fukui2009, Gubernatis2016}.

Although the above formulation should work theoretically, we sometimes encounter oscillatory behavior near the solution in practice.
This is because the Jacobian matrix \(\grad{\bm{f}}\) has large off-diagonal components, which makes the convergence slow.
To solve this problem and to speed up the convergence, we transform \cref{eq:tuneb_naive,eq:tuneg_naive} into the following mathematically equivalent form by taking their linear combinations:
\begin{gather}
    S\qty(\frac{2\pi}{L},0) - S\qty(0,\frac{2\pi}{K}) = 0, \label{eq:tuneb}                                                      \\
    S\qty(0,0) - \frac{1 + 4 \pi^2 R^2}{2} \qty[S\qty(\frac{2\pi}{L},0) + S\qty(0,\frac{2\pi}{K})] = 0. \label{eq:tuneg}
\end{gather}
The above transformation is based on the following physical insights;
first, \cref{eq:tuneb} is interpreted as the condition of isotropic aspect ratio.
As \(K\) is another length scale apart from \(L\), it is natural to expect that \cref{eq:tuneb} is more affected by \(K\) rather than \(\Gamma\).
In contrast, \cref{eq:tuneg} is the condition for the \textit{averaged} correlation length, which seems to be more sensitive to \(\Gamma\) being the control parameter of disorder.
Thus, by tuning \(K\) by \cref{eq:tuneb} and \(\Gamma\) by \cref{eq:tuneg}, we can expect that \(\grad{\bm{f}}\) is nearly diagonal.

\subsubsection{\texorpdfstring{Stochastic optimization of \(R\)}{Stochastic optimization of R}}
Suppose, with stochastic approximation or similar, we succeeded in tuning \(K\) to satisfy \cref{eq:tuneb}.
For now, \(\Gamma\) is still a free parameter.
From \cref{eq:fssx,eq:fsst}, we get \(\xi_x\qty(L,g) / L = \xi_\tau\qty(L,g) / K \sim \Xi\qty(L^{1 / \nu} g)\) for each \(\Gamma\).
As \(\Gamma\) makes the system disordered, we should observe a monotonically decreasing curve for scaled \(\xi_\ast\) near \(\Gamma_\text{c}\).
The gradient near \(\Gamma_\text{c}\) becomes steeper as \(\sim L^{1 / \nu}\), guaranteeing asymptotic convergence \(\Gamma - \Gamma_\text{c} \sim L^{- 1 / \nu}\) for every decent \(R\).
While the convergence is guaranteed for any positive \(R\), there exists an optimal value which achieves the fastest convergence to \(\Gamma_\text{c}\): the crossing point of the curves located at \(\Gamma = \Gamma_\text{c}\).
With this \(R\), \(\Gamma\) sticks to \(\Gamma_\text{c}\) for all \(L\) up to correction to scaling.
In terms of efficiency and correction reduction, it is strongly desirable to use this optimal value even though the above framework works for any \(R\).

We introduce an additional stochastic approximation to find this optimal \(R\).
First, we prepare two different system sizes, \(L\) and \(L' > L\).
The auxiliary system size \(L'\) is chosen so that \(\qty(L' / L)^{1 / \nu}\) is distinguishable from unity, say \(\qty(L' / L)^{1 / \nu} = 2\).
These two systems are assumed to be in the canonical ensembles at \(\qty(K,\Gamma)\) and \(\qty(K',\Gamma)\), respectively.
Next, we impose the following three conditions:
\begin{gather}
    S_L\qty(\frac{2\pi}{L},0) - S_L\qty(0,\frac{2\pi}{K}) = 0, \label{eq:tuneb1}                                                      \\
    S_{L'}\qty(\frac{2\pi}{L'},0) - S_{L'}\qty(0,\frac{2\pi}{K'}) = 0, \label{eq:tuneb2}                                                      \\
    \begin{multlined}
        S_{L'}\qty(0,0) \qty[S_L\qty(\frac{2\pi}{L},0) + S_L\qty(0,\frac{2\pi}{K})] \\
        - S_L\qty(0,0) \qty[S_{L'}\qty(\frac{2\pi}{L'},0) + S_{L'}\qty(0,\frac{2\pi}{K'})] = 0. \label{eq:tuneg_rev}
    \end{multlined}
\end{gather}
Here, we add subscripts to \(S\) to distinguish the quantities for different system sizes.
It is easy to see that these formulae are equivalent to \cref{eq:tuneb,eq:tuneg} for \(L\) and \(L'\), with \(R\) erased.
The point is that two systems are sharing \(R\), which is eventually eliminated, as well as \(\Gamma\); it is impossible to satisfy three constraints \cref{eq:tuneb1,eq:tuneb2,eq:tuneg_rev} simultaneously unless \(R\) is the optimal value.
We use \cref{eq:tuneb1,eq:tuneb2,eq:tuneg_rev} for optimizing \(K\), \(K'\), and \(\Gamma\), respectively, by which we can expect that \(\boldsymbol{A}\) is nearly diagonal, as discussed in the previous section.
Note that the left-hand side of \cref{eq:tuneg_rev} is an unbiased estimator due to statistical independence of \(S_L\) and \(S_{L'}\).

\subsection{Further details of computation and analysis}
\subsubsection{Choice of parameters}
To cover wide range of \(\sigma\), we investigate \(\sigma = 0.2,0.4,0.6,\ldots,2.4\) and two predicted universality boundaries, \(\sigma = 2 / 3\) and \(7 / 4\).
As for system size \(L\), we choose \(L = \lfloor \sqrt{2}^i + 1 / 2 \rfloor\) (rounding-off) for \(i = 6,7,\ldots\). The largest \(L\) for each \(\sigma\) is given in \cref{tab:data}.
The secondary system size \(L'\) is chosen to roughly satisfy \((L' / L)^{1 / \nu} \simeq 2\); \(L ' / L\) is \(16\) for \(\sigma = 0.2\), \(4\) for \(\sigma = 0.4,0.6\), \(3\) for \(\sigma = 2 / 3,0.8,\ldots,1.8\), and \(2\) for \(\sigma = 2.0,2.2,2.4\).
Along the chain direction, we impose periodic boundary conditions and take all the interactions from infinitely replicated mirror images into account to reduce the finite-size effect; we replace the coupling constant \(J_{ij} = 1/\qty|i - j|^{1 + \sigma}\) by
\begin{equation}
    \begin{split}
        J_{ij} & = \sum_{n = -\infty}^{\infty} \frac{1}{\qty|i - j + nL|^{1 + \sigma}}                                                          \\
               & = \frac{1}{L^{1 + \sigma}}\qty[\zeta\qty(1 + \sigma,\frac{\qty|i - j|}{L}) + \zeta\qty(1 + \sigma,1 - \frac{\qty|i - j|}{L})],
    \end{split}
    \label{eq:cint}
\end{equation}
where \(\zeta\qty(s,q)\) is the Hurwitz zeta function, \(\zeta\qty(s,q)= \sum_{n = 0}^{\infty} \qty(q + n)^{-s}\)~\cite{Nagle1970, Koziol2021, Luijten1997}.
In the simulations, \cref{eq:cint} is directly evaluated, and thus, there is no truncation or approximation employed.

For initial value of \(\Gamma\), we use \(\Gamma_\text{c}\) obtained by precalculation with \(L = 8\).
Unlike \(\Gamma\), on the other hand, \(K\) and \(K'\) are not so sensitive to initial values, and we thus use \(K = K' = 1\) for all \(\sigma\).
As discussed in \cref{sec:tuning}, the denominator \(n\) of \cref{alg:sa} is replaced by \(\lceil n / n_{\mathrm{block}} \rceil\) with \(n_{\mathrm{block}} = 10\).
Before starting parameter tuning, we run \(1,000\) MCS for equilibration.
Each simulation consists of \(10,000\) steps for parameter tuning and \(5,000\) for data collection.
At each tuning step, we first update configuration \(10\) MCS before collecting data for successive \(10\) MCS\@.
For better precision and stability, simulation is repeated \(\simeq 10^2\) times for each \(\sigma\) and \(L\).

\subsubsection{Post processing}
In ordinary MCMC simulations, data analysis is typically performed by binning and fitting to the FSS ansatz.
As for our method, the procedure is slightly different: binning, exponent estimation, and extrapolation to \(L \to \infty\).
First, we bin the sequence to estimate the mean and standard error as usual.
Next, instead of fitting with the whole data, we first estimate exponents for each parameter.
In this step, thanks to parameter tuning, we can rely on the quotient method~\cite{Ballesteros1996}, which brings us excellent stability and extensibility.
Recall that we have two sequences of a quantity \(Q_\ast\) with different system sizes.
According to the design of our method, these two observables \(Q_L\) and \(Q_{L'}\) are expected to behave as \(Q_L \sim L^{-x_Q}\) and \(Q_{L'} \sim L'^{-x_Q}\), respectively.
With this property, we can estimate \(x_Q\) just by taking the ratio of two \(Q_\ast\) as \(Q_{L'} / Q_L = \qty(L' / L)^{-x_Q}\) using binning and Jackknife resampling.
In our simulation with \(L' / L = \textrm{const}\)., the final estimation reduces to extrapolation of the estimated values to \(L \to \infty\), typically more stable than fitting to the FSS ansatz.
In this step, we are again empowered by the quotient method; we can relate the extrapolation function to scaling correction.
For example, correction-to-scaling form \(Q_L \simeq C L^{-x_Q} \qty(1 + A L^{- \upomega})\) naturally leads to \(Q_{L'} / Q_L \simeq \qty(L' / L)^{-x_Q} + A' L^{- \upomega}\), and multiplicative logarithmic corrections \(Q_L \simeq C L^{-x_Q} \log^p(L / L_0)\) does to \(Q_{L'} / Q_L \simeq \qty(L' / L)^{-x_Q} \qty[1 + \log(L' / L) / \log(L / L_0)]^p\).
For the critical point \(\Gamma_\text{c}\) and the optimal \(R\), we simply use corresponding scaling forms: \(\Gamma - \Gamma_\text{c} \sim L^{- \uptheta}\) and \(R \sim \textrm{const}\).

In the final extrapolation step, we fit all the possible sets of consecutive \(L\)'s to incorporate systematic errors.
This analysis is motivated by the fact that we often see difficulties for both large and small \(L\) due to ill-convergence and finite-size artifacts.
As there is no prior knowledge about reliable ranges of \(L\), we consider all the possible sets.
The final result is obtained by taking the weighted average of all the results, where the inverse square of the fitting error gives the weight.
For the associated error, we use the standard deviation of the whole data \textit{without} divided by the square root of the data size, treating them not as samples from the same distribution but as merely observed values.
For \(\Gamma_\text{c}\), as the data variance is sometimes below fitting uncertainties, we adopt the maximum of the latter as the final error.

Before closing this section, we should mention that the value of \(1/\nu\) cannot be estimated within the present analysis framework.
In the conventional finite-size scaling, \(1/\nu\) is estimated using data evaluated at \(\Gamma\) slightly shifted from the critical point \(\Gamma_{\text c}\).
However, in the present method, \(\Gamma\) is always automatically set to the finite-size critical point, so \(1/\nu\) cannot be evaluated.
Although it is still possible to estimate \(1/\nu\) separately combining the highly accurate values of \(\Gamma_\text{c}\), \(z\), and \(\upgamma/\nu\), etc, estimated by the present method, together with simulation results at off-criticality, but this is outside the scope of the present study.


\section{Result}\label{sec:result}
\subsection{Critical point}
\begin{figure*}[tbp]
    \centering
    \includegraphics[width=\linewidth]{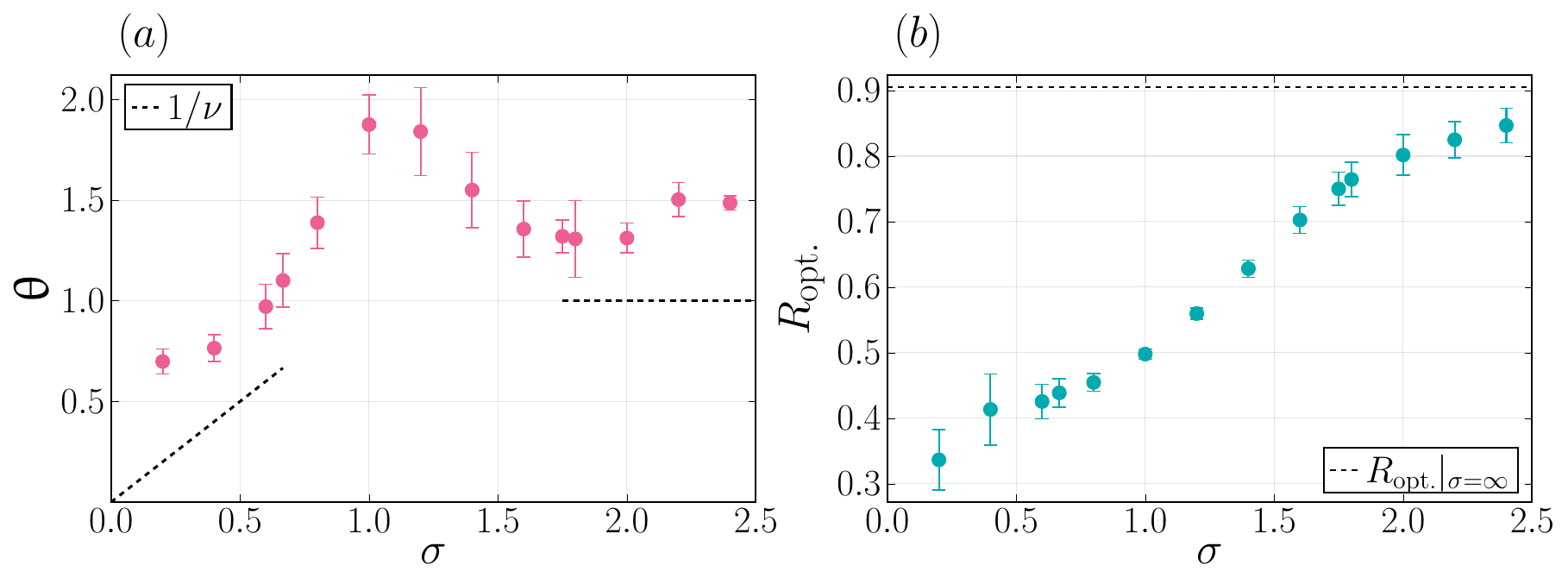}
    \caption{%
        (a)~Correction-to-scaling exponent estimated from \(\Gamma - \Gamma_\text{c} \sim L^{-\uptheta}\).
        (b)~Optimal \(R\) automatically determined during simulations.
        The horizontal dashed line shows the asymptotic value \(\simeq 0.905\)~\cite{Salas2000} in the \(\sigma = \infty\) limit.
    }\label{fig:value}
\end{figure*}

Our results for \(\Gamma_\text{c}\) are summarized in \cref{fig:diagram} and \cref{tab:data}~\cite{TFIData}.
By its design, our method is well suited for estimating \(\Gamma_\text{c}\), typically reaching \(\simeq 0.1 \%\) precision.
It also accurately reproduces the exact asymptotic behavior, \(\Gamma_\text{c} \to 1\), for \(\sigma \to \infty\)~\cite{Gubernatis2016, Lieb1961}.
In the opposite limit (\(\sigma \to 0\)), \(\Gamma_\text{c}\) increases proportionally to \(\sigma^{-1}\).
This behavior is intuitive since \(J = \sum_{\Delta > 0} J_{i,i + \Delta} \simeq \zeta\qty(1 + \sigma) \sim \sigma^{-1}\) for \(\sigma \ll 1\)~\cite{Whittaker1996, Jaschke2017}.
We find that the \(\sigma\)-dependence of \(\Gamma_\text{c}\) can be fitted quite well using the following empirical formula:
\begin{align}
    \Gamma_\text{c}(\sigma) = \frac{a}{\sigma^p + b \sigma} + 1
    \label{eq:empirical}
\end{align}
with \(a = 6.10(1)\), \(b = 2.994(8)\), and \(p = 3.087(2)\).
\Cref{eq:empirical} effectively captures the expected asymptotic behavior for for both \(\sigma \to 0\) and \(\sigma \to \infty\), and consistently describes the present numerical results within 95\% confidence interval across the entire range of \(\sigma\).

Additionally, as a byproduct of \(\Gamma_\text{c}\) estimation, we can extract the convergence rate exponent \(\uptheta\) from \(\Gamma - \Gamma_\text{c} \sim L^{-\uptheta}\) as shown in \cref{fig:value}{(a)}.
It is noteworthy that throughout the range, \(\uptheta\) does not become very small and seems to be always larger than \(1 / \nu\), which would appear here in ordinary FSS analysis.
This demonstrates that our \(R\)-tuning scheme accelerates convergence to the critical point.
For the optimal \(R\), we can see a monotonically increasing trend in \cref{fig:value}{(b)}.
For \(\sigma \gg 1\), the optimal \(R\) approaches \(R \simeq 0.905\), the previous estimate for the 2D classical Ising model with isotropic geometry~\cite{Salas2000}, justifying our method by imposing isotropy in the space-time plane.

\subsection{Critical exponents}
\subsubsection{Extrapolation with power-law corrections}
\begin{figure*}[tbp]
    \centering
    \includegraphics[width = \linewidth]{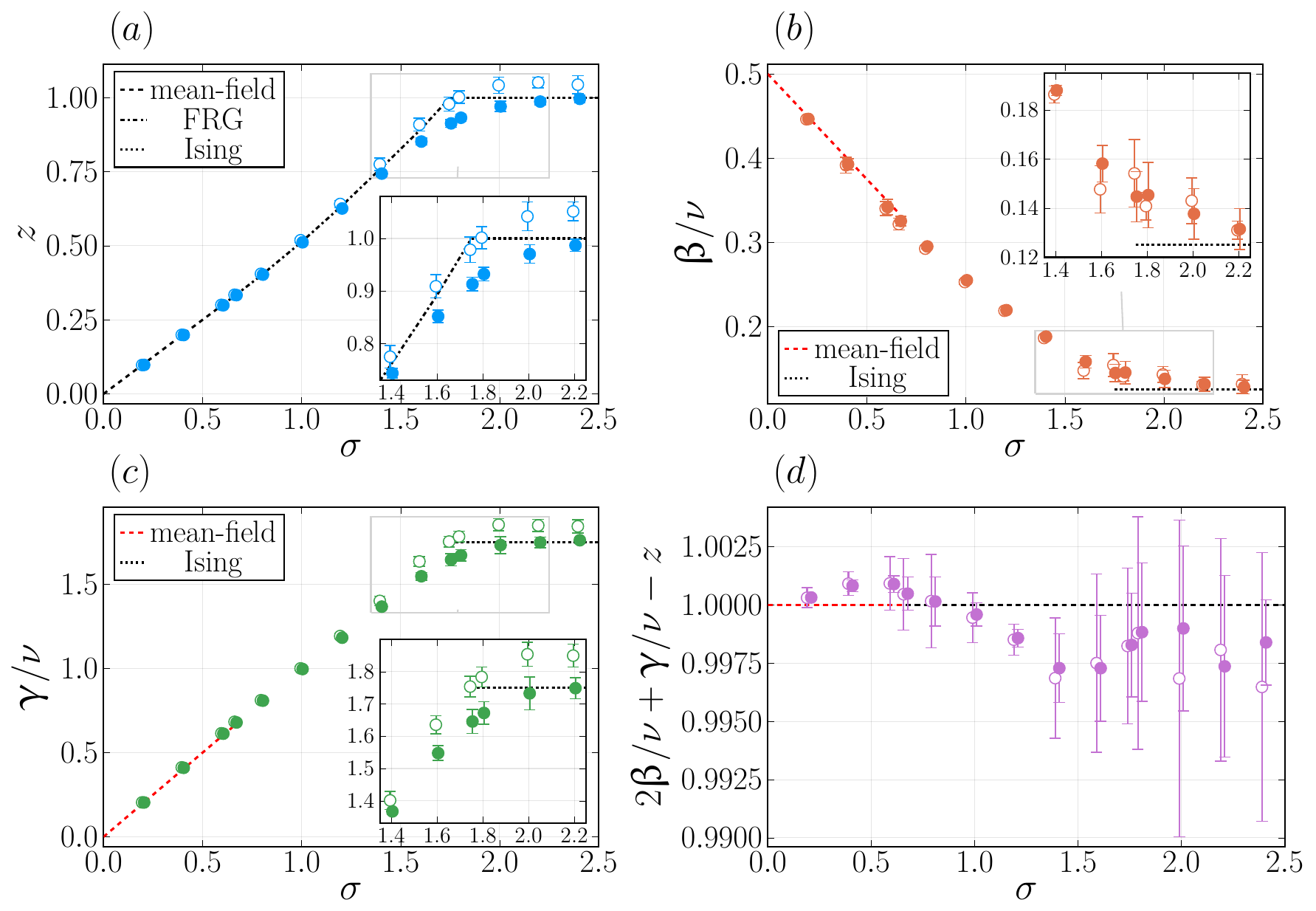}
    \caption{%
        Critical exponents estimated from power-law (filled circles) and logarithmic (open circles) extrapolations.
        The power-law and logarithmic extrapolation results are slightly shifted to the right and left, respectively, to avoid overlap.
        (a)~Dynamical exponent \(z\) estimated from the \(L\)-dependence of \(K\).
        For \(2 / 3 \leq \sigma \leq 7 / 4\), the functional RG prediction~\cite{Defenu2017} is plotted by the dash-dotted line.
        (b)~\(\upbeta / \nu\) estimated from \(m^2\).
        (c)~\(\upgamma / \nu\) estimated from \(S(0,\ 0)\).
        (d)~\(2\upbeta / \nu + \upgamma / \nu - z\), which should be equal to \(D = 1\) if the hyperscaling relation holds.
    }\label{fig:exp}
\end{figure*}

\Cref{fig:exp} summarizes our numerical results (filled circles) overlayed on theoretical predictions, obtained by power-law extrapolation assuming \(Q_{L'} / Q_L \simeq \qty(L' / L)^{-x_Q} + A' L^{- \upomega}\).
Before going into details, let us tell an essential caveat about \cref{fig:exp}.
As seen in three figures except \cref{fig:exp}{(a)}, theoretical results below \(\sigma = 2 / 3\) are drawn as dashed red lines.
This is because, in this region, the free energy density \(f\) is affected by dangerously irrelevant variable \(u\) (see \cref{sec:fss}).
Thus, we cannot see direct correspondence between \(L\)-dependence of observables and the critical exponents.
If one still wants to explore the region, as discussed in \citesref{Koziol2021, Binder1985, Flores-Sola2015}, extra care is needed to eliminate the effect of \(u\).
In this paper, we do not perform that kind of analysis for two reasons: The most crucial quantity \(z\) is still measurable in high precision, and our method is intentionally designed to cancel \(L^{1 / \nu}\) correction in \(\Gamma - \Gamma_\text{c}\), which is in principle not suitable for \(\nu\)-based analysis.
We do not necessarily expect data points to be consistent with the theoretical prediction in this region.

For \(\sigma \le 2 / 3\), \(z\) is quite close to the theoretical prediction, \(z=\sigma / 2\), as seen in \cref{fig:exp}{(a)}.
Interestingly, \(z\) seems not to be affected by severe multiplicative logarithmic corrections~\cite{Adelhardt2020, Dutta2001}, implying that we could bypass the problem by starting from correlation lengths as the fundamental quantities.
Along with the dangerously irrelevant variable problem, the correlation length likely obeys a qualitatively different scaling law compared to the free energy.
Unexpectedly, \(\upbeta / \nu\) and \(\upgamma / \nu\) are seemingly consistent with the theoretical prediction.
Next, for the nontrivial \(2 / 3 \le \sigma \le 7 / 4\), exponents monotonically change as pointed out in \citesref{Defenu2017, Koziol2021, Puebla2019, Sperstad2012}.
There are no discernable anomalies near \(\sigma = 1\), even though this point is predicted to be a tricritical point~\cite{Dutta2001, Anderson1971, Cannas1995, Luijten2001, Sandvik2003, Fukui2009, Humeniuk2020}, where the finite-temperature Kosterlitz-Thouless transition line touches the zero temperature.
As for the values, while \(z\) is still consistent with the theoretical prediction based on functional RG~\cite{Defenu2017} for \(\sigma = 0.8,\ 1.0\), it slowly deviates as \(\sigma\) increases, and at \(\sigma = 7 / 4\), it is about \(8\) \% smaller than the functional RG prediction.
Additionally, it is noteworthy that error bars are getting larger near \(\sigma = 7 / 4\).
From our convention for estimating errors for exponents, a larger error bar indicates that the extrapolated value severely depends on how we choose consecutive \(L\)'s to fit.
This observation could signify that we are trying to perform power-law fitting to the data that obey a completely different form.
To argue further, we will try another analysis in the next section.
Finally, for \(\sigma \ge 7 / 4\), we are over/underestimating exponents, though the discrepancy becomes smaller as \(\sigma\) goes away from \(7 / 4\).
We verify that the exponents become closer to the theoretical prediction outside the range before reaching the nearest-neighbor transverse-field Ising model at \(\sigma = \infty\).
Like the case of \(\sigma < 7 / 4\), error bars again get smaller as \(\sigma\) goes away from 7/4.
Note that the hyperscaling relation \(2\upbeta / \nu + \upgamma / \nu - z = 1\) is not seriously violated for all \(\sigma\), even though each term, \(\upbeta / \nu\), \(\upgamma / \nu\), or \(z\), is somtimes off from the theoretical prediction.

For the full data, see \cref{tab:data} (\(z\) only) and \citeref{TFIData}.

\subsubsection{Extrapolation with logarithmic corrections}
For now, while for small/large \(\sigma\), we have achieved satisfactory accuracy, we still have some room for improvement, especially near \(\sigma = 7 / 4\).
In this section, we will try another extrapolation scheme, \(Q_{L'} / Q_L \simeq \qty(L' / L)^{-x_Q} \qty[1 + \log(L' / L) / \log(L / L_0)]^p\), where we assume logarithmic corrections.
As a motivation for this scheme, first, let us recall that at \(\sigma = 2 / 3\), one can prove the existence of multiplicative logarithmic corrections~\cite{Adelhardt2020, Dutta2001}.
It is caused by the vanishing scaling exponent at the universality boundary, and thus, we would have logarithmic corrections at other universality boundaries.
Although this form of extrapolation is considered unsuitable except near \(\sigma = 2 / 3,\ 7 / 4\), we apply it for completeness.

The results obtained assuming the logarithmic corrections are plotted by open circles in \cref{fig:exp}.
We still have reasonable accuracy for all the exponents for \(\sigma \le 2 / 3\).
As a possible explanation for this precision, we note that in this range of \(\sigma\), our estimates at finite \(L\) are already very close to certain constants, even without extrapolation.
For such a case, any extrapolation would work well.
We have different trends from the previous results for \(2 / 3 \le \sigma \le 7 / 4\).
As for \(z\), we now have higher consistency with \citeref{Defenu2017} for all \(\sigma\).
The same is also seen in \(\upbeta / \nu\) and \(\upgamma / \nu\).
Finally, for \(\sigma \ge 7 / 4\), in contrast to the non-universal region, we have a larger error bar than the previous analysis.
The most severe problem is that the gap between our estimate and theory is not closing, signaling we are forcibly fitting with the incorrect form.
Again, the hyperscaling relation is not seriously violated for all \(\sigma\).

For the full data, see \cref{tab:data} (\(z\) only) and \citeref{TFIData}.

\subsubsection{Univesarity boundary}
While we achieve excellent accuracy for small \(\sigma\), we still suffer from strong finite-size effects for larger \(\sigma\).
The true \(\sigma\)-dependence of the critical exponents is expected to show cusp-like behavior at the universality boundary between the non-universal and 2D Ising universality classes, as indicated by the dotted and dash-dotted lines in \cref{fig:exp}{(a)}.
However, due to strong finite-size corrections near the boundary, the numerical results exhibit continuous changes, with the cusp being smoothed out.
In fact, the power-law extrapolation results for \(z\) and \(\upgamma/\nu\) show gradual increases even for \(\sigma \ge 2\), where it is certain that the 2D Ising universality is realized.
From this observation, we expect that the power-law extrapolation gives the lower bound for true \(z\) and \(\upgamma/\nu\) in the vast region spanning from the non-universal to the 2D Ising universality classes.

\begin{figure}[tbp]
    \centering
    \includegraphics[width=\linewidth]{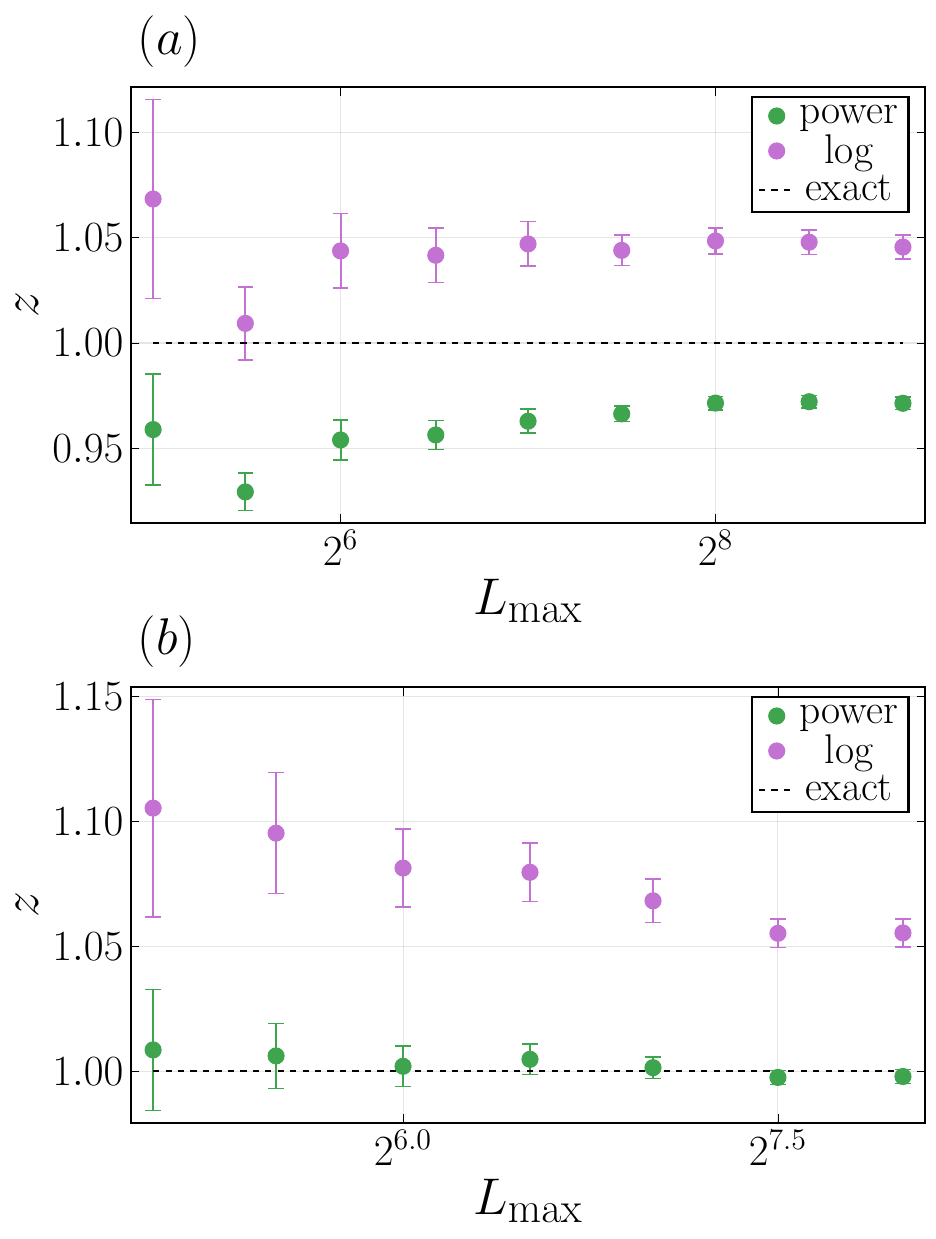}
    \caption{%
        Extrapolation of \(z\) at \(\sigma = 2.0\) (a) and \(\sigma = 2.4\) (b). Here, we change \(L_{\max}\) with fixing \(L_{\min} = 8\).
        Theoretical prediction is \(z = 1\) for both cases.
        As seen, especially when far from the exact value, power-law extrapolation has increasing trends, while logarithmic extrapolation does the opposite.
    }\label{fig:conv}
\end{figure}

On the other hand, the logarithmic extrapolation should work best at the universality boundary but is thought to give poorer estimates than power-law extrapolation in regions away from there.
Below, we compare the behavior of both in more detail at \(\sigma=2\) and \(2.4\).
In \cref{fig:conv}, we plot \(z\) estimated from various \(L_{\max}\) for \(\sigma = 2.0\) and \(\sigma = 2.4\) with fixing \(L_{\min} = 8\).
The power-law and the logarithmic extrapolations have opposite trends; the former approaches from below, while the latter from above.
This trend is more apparent when the discrepancy is significant, as in \cref{fig:conv}.
Thus, we can expect that the true value lies between the two extrapolated values, and the gap would close if we could reach sufficiently large \(L\).

Generally speaking, the two analyses have complementary roles.
The power-law and logarithmic extraporations give lower and upper bounds, respectively.
However, just at the universal boundary, the logarithmic extrapolation should give much more accurate results than the power-law extrapolation, as the former takes into account the effects of vanishing scaling exponents precisely.

Based on the above discussions, we would say that the universality boundary can be located by identifying \(\sigma\) where the logarithmic extrapolation gives \(z=1\).
Thus, from the present numerical data, we conclude that the universality boundary is not at \(\sigma = 2\), but at \(\sigma = 7 / 4\).
The same conclusion is also drawn from the logarithmic extrapolation results for \(\upgamma/\nu\) (\cref{fig:exp}{(c)}).
Our conclusion supports the theoretical prediction given in~\cite{Defenu2017}.
As for \(\upbeta/\nu\), the logarithmic extrapolation has quite large error bar near the universality boundary, and both of \(\sigma = 7/4\) and \(\sigma = 2\) are supported if we consider 95\% confidence interval.

\begin{table}
    \begin{ruledtabular}
        \begin{tabular}{lS[table-format=2.7]S[table-format=1.7]S[table-format=1.6]r}
            {\(\sigma\)} & {\(\Gamma_\text{c}\)} & {\(z\) (pow)} & {\(z\) (log)} & {\(L_{\max}\)} \\ \hline 
            0.2          & 11.08(9)              & 0.098(2)      & 0.098(2)      & 128            \\
            0.4          & 5.85(2)               & 0.199(2)      & 0.199(3)      & 1448           \\
            0.6          & 4.043(6)              & 0.299(2)      & 0.300(3)      & 1024           \\
            \(2 / 3\)    & 3.673(2)              & 0.3334(7)     & 0.335(3)      & 1024           \\
            0.8          & 3.1041(8)             & 0.403(2)      & 0.405(4)      & 512            \\
            1.0          & 2.5264(4)             & 0.512(2)      & 0.519(8)      & 512            \\
            1.2          & 2.1396(4)             & 0.626(3)      & 0.64(2)       & 512            \\
            1.4          & 1.8691(4)             & 0.744(9)      & 0.78(3)       & 512            \\
            1.6          & 1.6731(3)             & 0.85(2)       & 0.91(3)       & 512            \\
            \(7 / 4\)    & 1.5609(3)             & 0.91(2)       & 0.98(3)       & 362            \\
            1.8          & 1.5288(6)             & 0.93(2)       & 1.00(3)       & 362            \\
            2.0          & 1.4208(2)             & 0.97(2)       & 1.04(3)       & 362            \\
            2.2          & 1.3389(1)             & 0.99(2)       & 1.05(2)       & 181            \\
            2.4          & 1.2755(2)             & 1.00(2)       & 1.04(4)       & 181            \\
        \end{tabular}
    \end{ruledtabular}
    \caption{%
        Estimates of \(\Gamma_\text{c}\) and \(z\) based on power-law (\(\Gamma_\text{c}\) and \(z\), first and second columns) and  logarithmic (\(z\), third column) extrapolations, along with the maximum \(L\) for simulations.
        The complete data is available in \citeref{TFIData}.
    }\label{tab:data}
\end{table}


\section{Discussion}\label{sec:discussion}
In the present paper, we analyze LRTFIM with a new approach that combines the order-N QMC with stochastic parameter optimization.
First, we conclude that the mean-field universality class for \(\sigma \le 2 / 3\).
Our proposed method is highly effective and consistent with analytical predictions.
Additionally, our scheme based on correlation lengths can bypass the problems caused by dangerously irrelevant variables.
For the non-universal regime, \(2 / 3 \le \sigma \le 7 / 4\), we observe monotonically varying exponents, which are consistent with previous studies~\cite{Defenu2017, Koziol2021, Puebla2019, Sperstad2012}.
The value of the dynamical exponent, \(z\), is consistent with theoretical prediction~\cite{Defenu2017}.
Still, for \(\sigma \gtrapprox 1.2\), we needed to adopt the logarithmic extrapolation scheme, not the power-law one that gives smaller estimates.
Additionally, from the \(\sigma\) dependence of \(z\), we conclude that the upper boundary of the non-universal region is \(\sigma = 7 / 4\).
In addition to the critical exponents, we estimate the critical points consistent with the previous studies~\cite{Zhu2018, Fey2016, Koziol2021}, but with much higher precision.
We also find that our method is capable of accelerating the convergence of finite-size \(\Gamma_\text{c}\) faster than the ordinary \(L^{-1 / \nu}\) scaling without any hand-tuning.
As a by-product of the convergence acceleration method, we also obtain the critical amplitude at the critical point: \(R\) for \(\xi_x / L\) and \(\xi_\tau / K\).

Next, let us assess our method.
Recall that our method consists of three components: MCMC sampler, automatic parameter tuning, and automatic hyperparameter tuning.
These components are mostly independent of the model subject to the analysis; we can apply our method to other models straightforwardly.
For example, we can directly apply it to the higher-dimensional LRTFIM by replacing the MCMC sampler.
We assume little about the model.
To be specific, we do not need to know \(\Gamma_\text{c}\), optimal \(R\), or even more subtle information such as the existence of dangerously irrelevant variables.
Our method has another advantage that the conventional methods do not have:
It automatically reduces finite-size artifacts.
Related to this, let us mention \citeref{Hasenbusch1999}, which can be regarded as a precursor of our method, though the proposed method is unsuitable for MCMC\@.

Finally, let us argue possible improvements and open questions.
For LRTFIM, finite-size artifacts are still prominent, and it is utterly essential to reduce them further.
For that purpose, it is worth trying to change the model detail itself, such as swapping the \(x\) and \(\tau\) axes, mixing short-range interaction for \(x\) direction, or even tuning the strength of long-range interaction~\cite{Sperstad2012, Hasenbusch1999}.
Additionally, the way for estimating the critical point is not the only one; for instance, extrema of binder cumulant~\cite{Sperstad2010} or ratio of partition function under different boundary conditions~\cite{Hasenbusch1999} can also serve as tuning subjects.
Although it may lead to smaller finite-size corrections, we regrettably do not have suitable unbiased estimators.
Ideally, we could change the imposed constraints depending on what we try to measure or use the subjected quantity as a tuning subject.
As for the algorithm itself, there is still room for improvement, especially regarding the convergence of \(\Gamma\).
To deal with it, zero-temperature MCMC simulation~\cite{Todo2006} may be helpful.
Actually, we can further reduce the number of parameters to \(1\) with this algorithm: for 1D optimization, there are plenty of stable algorithms, even for ill-conditioned problems.

As closing remarks, let us locate our work.
Our method, which automatically guides users to the critical point faster than the standard methods, will free numerical simulations from the burden of exhausting try-and-error.
Additionally, as one of the many studies of LRTFIM, our results would be regarded as a piece of the puzzle illustrating the power and limitation of direct numerical simulation approaches.


\appendix

\begin{acknowledgments}
    The authors thank Tsuyoshi Okubo and Hidemaro Suwa for fruitful discussions and comments.
    This work was supported by the Center of Innovation for Sustainable Quantum AI, JST Grant Number JPMJPF2221, and JSPS KAKENHI Grant Numbers JP20H01824 and JP24K00543.
    SS was supported by Grant-in-Aid for JSPS Fellows Grant Number JP24KJ0890.
    The computation in this paper was done partly by the Nekoya/AI cluster at the Institute for Physics of Intelligence, the University of Tokyo, and the facilities of the Supercomputer Center, the Institute for Solid State Physics, the University of Tokyo.
    SS acknowledges support from FoPM, WINGS Program, the University of Tokyo.
\end{acknowledgments}

\section{Dealing with nonmonotonicity}\label{sec:nonmonotonicity}
\begin{figure}[tbp]
    \centering
    \includegraphics[width=\linewidth]{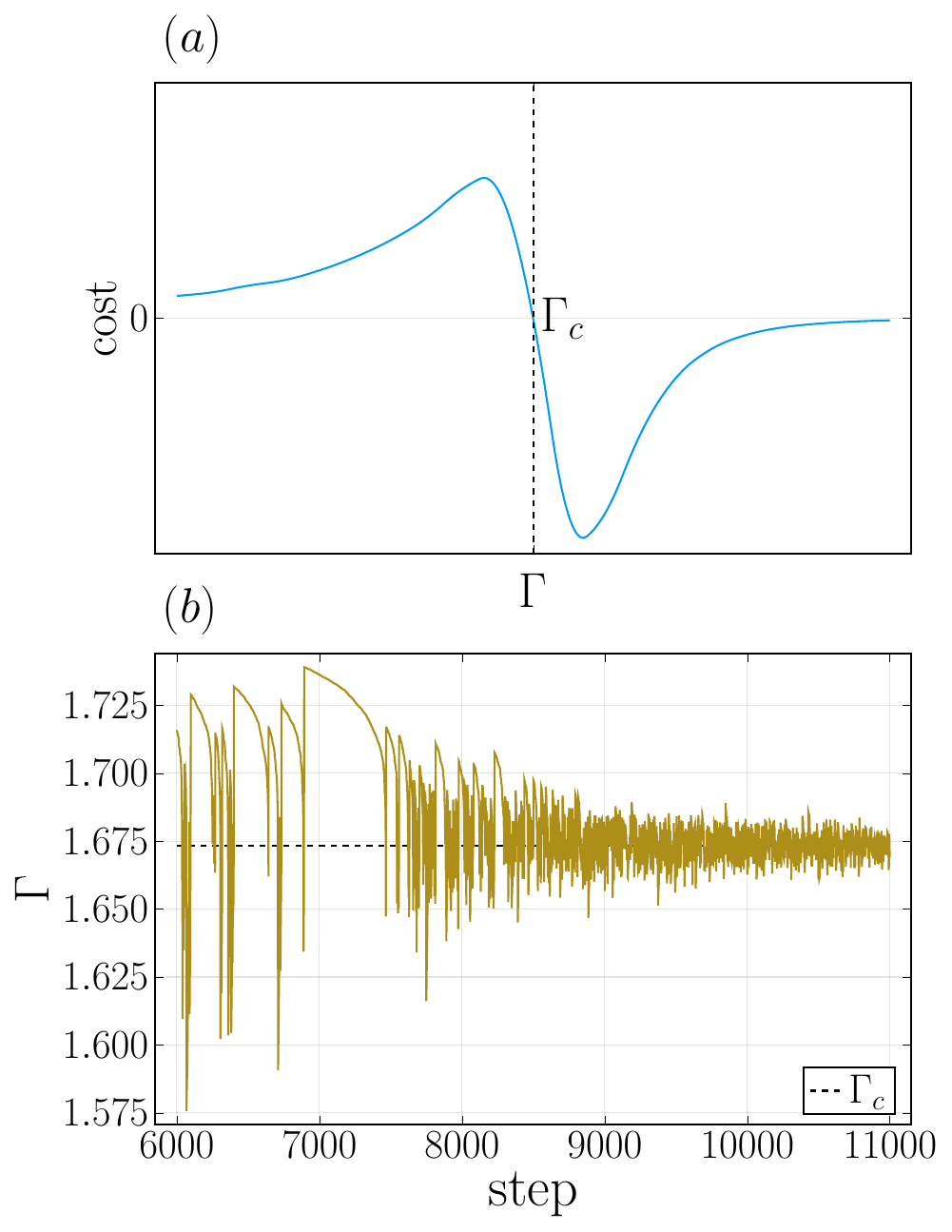}
    \caption{%
        (a)~Qualitative behavior of the cost function of \(\Gamma\) [\cref{eq:tuneg_rev}].
        Its strong nonmonotonicity results in a severe convergence problem if \(\boldsymbol{A}\) is too large.
        (b)~An example trajectory of \(\Gamma\) tuned based on \cref{eq:tuneg_rev}.
        Before reaching \(\Gamma_\text{c}\), it periodically jumps over \(\Gamma_\text{c}\) from below and then takes much time to come back.}\label{fig:volatile}
\end{figure}

For a monotonic function \(f\), it is a reasonable strategy to choose very large \(A\) in \cref{alg:sa}: convergence rate is limitlessly improved by increasing \(A\), whereas asymptotic fluctuation becomes larger~\cite{Yasuda2015}.
However, especially for \(\Gamma\) tuning with \cref{eq:tuneg_rev} it does not apply.
From here, as \(K\) and \(K'\) rarely suffer from ill-convergence, we assume that only \(\Gamma\) is off from the target value, and \(K,\ K'\) instantly follow the change of \(\Gamma\).
Under this assumption, we can directly plot the cost function of \(\Gamma\) as shown in \cref{fig:volatile}{(a)} by toggling off \(\Gamma\) tuning.
Obviously, it is not monotonic; it has maximum and minimum points near \(\Gamma_\text{c}\) and rapidly decays away from the critical region.
Near \(\Gamma_\text{c}\), its asymptotic behavior is given by \((\Gamma - \Gamma_\text{c}) L^{1 / \nu}\) as usual.
Similarly, the gradient of the cost function also has the same scaling; the amplitudes at two extrema are \(\order{1}\) as the result of adaptive cost normalization (see \cref{alg:cycle}).
In addition, to address the source of problems, we also need to consider the asymptotic behavior for \(\Gamma \to 0\) and \(\Gamma \to \infty\).
In the former limit, we have the classical long-range Ising model with two different parameter sets \(\qty(L,\ K)\) and \(\qty(L',\ K')\).
As it is far below the critical point, they are in their ordered phases.
Thus, using \cref{eq:dsf}, the cost function is evaluated to be \(0\) unless \(q\) and \(\omega\) are both zero.
We have a vanishing correlation in the opposite limit due to strong quantum fluctuations induced by large \(\Gamma\).
Therefore, we get \(0\) in this limit again.
To sum up, the cost function rapidly becomes zero for both ends and has \(\order{1}\) amplitude only a very narrow range around \(\Gamma_\text{c}\).
With this shape of the cost function, we encounter a severe problem when approaching \(\Gamma_\text{c}\) from above; stochastic approximation takes a long time to escape from the flat region located at \(\Gamma \gg \Gamma_\text{c}\), and once it hits the minimum, \(\Gamma\) experiences a big jump proportional to \(\boldsymbol{A}\).
After crossing \(\Gamma_\text{c}\), a similar jump is attempted at the maximum for the opposite direction.
Obviously, if \(\boldsymbol{A}\) is too large, it lands on the negative region far above \(\Gamma_\text{c}\).
Consequently, we sometimes observe jump-and-stick behavior as shown in \cref{fig:volatile}{(b)}.
To avoid this, we must make \(\Gamma\) restricted between two peaks.
Based on these observations, we adopt a precise estimate of \(\Gamma_\text{c}\) as the initial value and set \(\boldsymbol{A}\) to be small enough.
For our scheme designed to negate \(L\)-dependence of \(\Gamma\), it suffices to run small precalculation to estimate \(\Gamma_\text{c}\) with \(L \simeq 10\).

\section{Heuristics for optimization}\label{sec:optimization}

\subsection{Aggressive tuning}\label{sec:tuning}
In realistic situations, we are more concerned about convergence than statistical error.
To incorporate this demand, we replace \(A / n\) with \(A / \lceil n / n_{\mathrm{block}} \rceil\) in \cref{alg:sa}.
With this replacement, because the denominator takes each integer value \(n_{\mathrm{block}}\) times before decrement, we can make the feedback more aggressive.
When feedback is so volatile that parameters become negative, we replace them with their absolute values.
This replacement is safe since it only occurs at the very beginning of the iteration.

\subsection{Enriching the quality of samples}
This rule applies only when random data come from MCMC\@.
Due to abrupt parameter changes caused by stochastic approximation, the system is slightly off from the true equilibrium~\cite{Weigel2021}.
Even though this bias problem is gradually reduced as the iteration proceeds, it is still preferable to run a few MCS for equilibration at each iteration.
Additionally, to prevent stochastic approximation from being too volatile, it is also advisable to feed the average of multiple samples to the optimizer.
This operation could also be related to the convergence in terms of reducing autocorrelation artifacts not included in the original algorithm~\cite{Robbins1951}.

\subsection{Choosing hyperparameters}
A major difficulty comes from the apparent nonuniformity of three parameters \(K\), \(K'\), and \(\Gamma\) and also their cost functions.
To make matters worse, the degrees of their anisotropy greatly depend on \(\sigma\), making it hard to manually find \(\boldsymbol{A}\) that works.
To cope with that, we propose a simple heuristic.
For each cost function, we compute \(\sqrt{\sum_{i = 1}^{n} f_i^2 / n}\), where \(f_i\) is the \(i\)-th sample.
This gives us a rough estimate of the cost function's standard deviation, and thus \(f_n / \sqrt{\sum_i f_i^2 / n}\) would have approximately unit variance.
Additionally, we also multiply it by the current parameter \(x_n\), so that \(x_n f_n / \sqrt{\sum_i f_i^2 / n}\) has relatively uniform variance for all parameters.
This simple trick reduces the number of hyperparameters to one, which uniformly controls the feedback amplitude for all parameters.
In our simulations, we use \(A = 2\) for \(L \leq 91\) and otherwise \(A = 1\), in view of the optimal \(A\) monotonically decreasing~\cite{Yasuda2015}.
The overall algorithm is summarized in \cref{alg:sa_norm}.
\begin{algorithm}[H]
    \begin{algorithmic}[1]
        \REQUIRE{Current estimate \(x_n\), hyperparameter \(A\)}
        \STATE{Sample \(f_n \sim f\qty(x_n) + \varepsilon\)}
        \STATE{\(x_{n + 1} \leftarrow x_n + A x_n f_n / \qty(\sqrt{n \sum_i f_i^2})\)}
        \ENSURE{\(x_{n + 1}\)}
    \end{algorithmic}
    \caption{Adaptive cost normalization.}\label{alg:sa_norm}
\end{algorithm}

\section{Detailed optimization procedure}

\begin{algorithm}[H]
    \begin{algorithmic}[1]
        \REQUIRE{\(\qty(K_n,\ K'_n,\ \Gamma_n)\)}
        \STATE{Equilibrate the systems}
        \STATE{Sample \(S_L\qty(0,\ 0),\ S_L\qty(2\pi / L,\ 0),\ S_L\qty(0,\ 2\pi / K_n)\)}
        \STATE{Sample \(S_{L'}\qty(0,\ 0),\ S_{L'}\qty(2\pi / L',\ 0),\ S_{L'}\qty(0,\ 2\pi / K'_n)\)}
        \STATE{Sample \(Q\qty(L,\ K_n,\ \Gamma_n),\ Q\qty(L',\ K'_n,\ \Gamma_n)\) (optional)}
        \STATE{\(f_n^K \leftarrow S_L\qty(2\pi / L,\ 0) - S_L\qty(0,\ 2\pi / K_n)\)}
        \STATE{\(f_n^{K'} \leftarrow S_{L'}\qty(2\pi / L',\ 0) - S_{L'}\qty(0,\ 2\pi / K'_n)\)}
        \STATE{\(f_n^{\Gamma} \leftarrow S_{L'}\qty(0,\ 0) \qty[S_L\qty(2\pi / L,\ 0) + S_L\qty(0,\ 2\pi / K_n)] - S_L\qty(0,\ 0) \qty[S_{L'}\qty(2\pi / L',\ 0) + S_{L'}\qty(0,\ 2\pi / K'_n)]\)}
        \STATE{\(f_n^{K} \leftarrow f_n^{K} / \sqrt{\sum_{i = 1}^{n} {f_i^{K}}^2 / n}\)}
        \STATE{\(f_n^{K'} \leftarrow f_n^{K'} / \sqrt{\sum_{i = 1}^{n} {f_i^{K'}}^2 / n}\)}
        \STATE{\(f_n^{\Gamma} \leftarrow f_n^{\Gamma} / \sqrt{\sum_{i = 1}^{n} {f_i^{\Gamma}}^2 / n}\)}
        \STATE{\(n_d \leftarrow \lceil n / n_{\mathrm{block}} \rceil\)}
        \STATE{\(K_{n + 1} \leftarrow K_n + A K_n f_n^{K} / n_d\)}
        \STATE{\(K'_{n + 1} \leftarrow K'_n + A K'_n f_n^{K'} / n_d\)}
        \STATE{\(\Gamma_{n + 1} \leftarrow \Gamma_n + A \Gamma_n f_n^{\Gamma} / n_d\)}
        \STATE{\(\qty(L,\ K_{n},\ \Gamma_{n}) \to \qty(L,\ K_{n + 1},\ \Gamma_{n + 1})\)}
        \STATE{Update \(\qty(L',\ K'_{n},\ \Gamma_{n}) \to \qty(L',\ K'_{n + 1},\ \Gamma_{n + 1})\)}
        \ENSURE{Update \(\qty(K_{n + 1},\ K'_{n + 1},\ \Gamma_{n + 1})\)}
    \end{algorithmic}
    \caption{Detailed procedure at each stochastic optimization step.}\label{alg:cycle}
\end{algorithm}

We give the detailed procedure at each stochastic optimization step in \cref{alg:cycle}.
Before starting data collection, we equilibrate the systems with several MCS\@.
After equilibration, the samplers are ready to generate data: \(S_L(0,\ 0)\), \(S_L(2\pi / L,\ 0)\), \(S_L(0,\ 2\pi / K_n)\), \(S_{L'}(0,\ 0)\), \(S_{L'}(2\pi / L',\ 0)\), \(S_{L'}(0,\ 2\pi / K'_n)\) and optionally \(Q_L\) and \(Q_{L'}\).
In our implementation, squared magnetization \(m^2 = (\sum_i Z_i / L)^2\) is sampled to estimate \(\upbeta / \nu\).
As previously mentioned, averages are taken over multiple samples to reduce fluctuation/autocorrelation artifacts.
Next, we compute cost functions \(\qty(f_n^{K},\ f_n^{K'},\ f_n^{\Gamma})\) from the sampled data and normalize them.
Finally, we update parameters \((K_n,\ K'_n,\ \Gamma_n)\) by the normalized cost functions.

\bibliography{ref}

\end{document}